\documentclass[prx, twocolumn,  superscriptaddress]{revtex4-2}
\usepackage{setspace}
\usepackage[table]{xcolor}
\usepackage{tabularx}\usepackage[colorlinks, citecolor=blue, linkcolor=blue, urlcolor=blue, linktocpage]{hyperref}

\usepackage{tikz}
\usepackage[utf8]{inputenc}
\usepackage{graphicx}
\usepackage{epstopdf}
\usepackage{amsmath}
\usepackage{mathtools}
\usepackage[]{natbib}

\usepackage{graphicx}
\setcitestyle{numbers}
\usepackage{eurosym}
\usepackage{tikz}
\usepackage{amsthm}

\usepackage{enumitem}
\setitemize{noitemsep,topsep=0pt,parsep=0pt,partopsep=0pt,leftmargin=*}
\usepackage{amssymb}

\renewcommand{\vec}[1]{\textbf{#1}}

\usepackage{mdframed}
\usepackage{enumitem}

\DeclareMathOperator*{\argmax}{arg\,max}

\def\z0{{z}_{t_0}}

\def\z{{z}}
\def\x{\mathbf{\chi}}

\def\th{\mathbf{\theta}}
\def\ph{\mathbf{\phi}}

\begin{document}

\author{Blake A. Wilson} 
\affiliation{Elmore Family School of Electrical and Computer Engineering and Purdue Quantum Science and Engineering Institute, Purdue University, West Lafayette, IN 47907, USA}
\affiliation{Quantum Science Center, Oak Ridge National Laboratory, Oak Ridge, TN 37831, USA}

\author{Jonathan Wurtz}
\affiliation{QuEra Computing Inc., 1284 Soldiers Field Road, Boston, MA, 02135, USA}

\author{Vahagn Mkhitaryan}
\affiliation{Elmore Family School of Electrical and Computer Engineering and Purdue Quantum Science and Engineering Institute, Purdue University, West Lafayette, IN 47907, USA}
\affiliation{Quantum Science Center, Oak Ridge National Laboratory, Oak Ridge, TN 37831, USA}

\author{Michael Bezick}
\affiliation{School of Computer Science, Purdue University, West Lafayette, IN 47907, USA}

\author{Sheng-Tao Wang}
\affiliation{QuEra Computing Inc., 1284 Soldiers Field Road, Boston, MA, 02135, USA}

\author{Sabre Kais}
\affiliation{Elmore Family School of Electrical and Computer Engineering and Purdue Quantum Science and Engineering Institute, Purdue University, West Lafayette, IN 47907, USA}
\affiliation{Quantum Science Center, Oak Ridge National Laboratory, Oak Ridge, TN 37831, USA}
\affiliation{Department of Chemistry, Purdue University, West Lafayette, IN 47907, USA}
\affiliation{School of Computer Science, Purdue University, West Lafayette, IN 47907, USA}

\author{Vladimir M. Shalaev}
\affiliation{Elmore Family School of Electrical and Computer Engineering and Purdue Quantum Science and Engineering Institute, Purdue University, West Lafayette, IN 47907, USA}
\affiliation{Quantum Science Center, Oak Ridge National Laboratory, Oak Ridge, TN 37831, USA}

\author{Alexandra Boltasseva}
\affiliation{Elmore Family School of Electrical and Computer Engineering and Purdue Quantum Science and Engineering Institute, Purdue University, West Lafayette, IN 47907, USA}
\affiliation{Quantum Science Center, Oak Ridge National Laboratory, Oak Ridge, TN 37831, USA}

\title{Non-native Quantum Generative Optimization with Adversarial Autoencoders}

\date{\today}
\begin{abstract}
Large-scale optimization problems are prevalent in several fields, including engineering, finance, and logistics.
However, most optimization problems cannot be efficiently encoded onto a physical system because the existing quantum samplers have too few qubits. Another typical limiting factor is that the optimization constraints are not compatible with the native cost Hamiltonian.
This work presents a new approach to address these challenges. We introduce the adversarial quantum autoencoder model (AQAM) that can be used to map large-scale optimization problems onto existing quantum samplers while simultaneously optimizing the problem through latent quantum-enhanced Boltzmann sampling.
  We demonstrate the AQAM on a neutral atom sampler, and showcase the model by optimizing $64\text{px} \times 64$px unit cells that represent a broad-angle filter metasurface applicable to improving the coherence of neutral atom devices. Using 12-atom simulations, we demonstrate that the AQAM achieves a lower Renyi divergence and a larger spectral gap when compared to classical Markov Chain Monte Carlo samplers. 
  Our work paves the way to more efficient mapping of conventional optimization problems into existing quantum samplers.
\end{abstract}
\maketitle

Quantum sampling algorithms \cite{Benedetti2019ParameterizedModels,Huang2021Information-theoreticLearningb, Wurtz2022IndustryProblems, Wurtz2024SolvingAlgorithms, King2021ScalingMagnets, Layden2023Quantum-enhancedCarlo, Nguyen2023QuantumArrays} use quantum dynamics to sample probability distributions over bitstrings that are often intractable to sample using classical computers, such as quantum phases \cite{Ebadi2021QuantumSimulator, Kairys2020SimulatingAnnealing, Cao2019QuantumComputing}, optimization landscapes \cite{Lucas2014IsingProblems, Wurtz2024SolvingAlgorithms, King2015BenchmarkingMetric}, and quantum-advantage experiments \cite{Huang2021Information-theoreticLearningb, Aaronson2010TheOptics, King2021ScalingMagnets, Coyle2020TheMachine}.
Similarly, some industry-focused sampling problems, such as circuit design \cite{Bian2016MappingDiagnosis}, graph coloring \cite{Nguyen2023QuantumArrays}, and traveling salesperson \cite{Tsukamoto2017AnProblems}, are \#P sampling problems \cite{Arora2009ComputationalApproach} and can be more efficiently encoded onto a quantum sampler because of their {\it native} reduction to the sampler's cost Hamiltonian \cite{Wurtz2024SolvingAlgorithms, Farhi2014AAlgorithm}.
Certain problems can be reduced to Quadratic Unconstrained Binary Optimization (QUBO) using their Ising formulations \cite{Lucas2014IsingProblems, Nguyen2023QuantumArrays}.
In turn, QUBO can easily be mapped onto a physical system of superconducting qubits \cite{Bian2016MappingDiagnosis, Lucas2014IsingProblems}, or Maximum Independent Set (MIS) on neutral atoms \cite{Nguyen2023QuantumArrays,Wurtz2022IndustryProblems}. The low-energy states of the reduced QUBO or MIS problem can then be sampled using available quantum hardware. After that, one can invert the reduction to generate the native sample bitstrings from the low-energy states.
However, several {\it non-native} sampling problems like material design \cite{Wilson2021MachineProblems, Wilson2021MetasurfaceSampling}, topology optimization \cite{Lin2019TopologyMetasurfaces, Kudyshev2020Machine-learning-assistedOptimization}, and quantum device design \cite{Ebadi2021QuantumSimulator, Preskill2018QuantumBeyond, Bauer2020QuantumScience} do not have efficient mappings from bitstrings to designs. Notably, finding new ways to efficiently sample these problems could have a wide-reaching impact on several industry applications.

\begin{figure}[!b]
    \includegraphics[width=0.8\columnwidth]{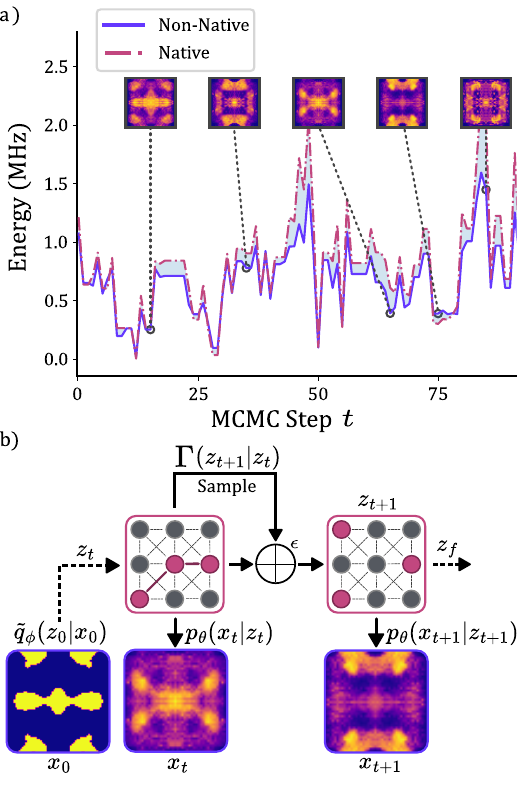}
  \caption{
  \textbf{ Adversarial Quantum Autoencoder Model (AQAM)}: a generative quantum machine learning method \cite{Kao2023ExploringChemistry, Lopez-Piqueres2023SymmetricOptimization} for MCMC sampling of non-native design spaces, shown here generating novel photonic metasurface unit cells. 
  (A) By matching the native energy for each native sample $\z_t$ (Purple line) with the objective function of the non-native design space (blue line), a native MCMC algorithm can explore a Boltzmann distribution in the non-native space (displayed images) by decoding samples from the native space.
  (B). A non-native unit cell design $\x$ (left) is encoded into a native bitstring $\z_0$ using a classical encoder $\tilde{q}_\phi(\z_0|\x)$. Each native bitstring sample $\z_t$ is updated to the next sample $\z_{t+1}$ using a quantum sampler $\z_{t+1} \sim \Gamma_3(\z_{t+1} \vert \z_{t})$.
  }
  \label{fig:latent_MCMC}
\end{figure}
This work presents a new, generic quantum machine learning method \cite{Perdomo-Ortiz2018OpportunitiesComputers, Vinci2019AAnnealers} to map non-native problems onto quantum samplers using an \textbf{adversarial quantum autoencoder model} (AQAM). The proposed AQAM (Figure \ref{fig:latent_MCMC}) uses a classical autoencoder \cite{Kingma2013Auto-EncodingBayes} trained to map bitstring measurement outcomes to and from novel candidate designs \cite{Rolfe2017DiscreteAutoencoders}. In this way, training an AQAM is a type-1 non-native hybrid algorithm where the mapping function is learned instead of defined \cite{Wurtz2024SolvingAlgorithms}. The ultimate goal of the algorithm is to generate a Boltzmann distribution over non-native designs given some objective $C_\chi(\chi)$ and inverse temperature $\beta$.
The quantum device then works as a co-processor \cite{Benedetti2019ParameterizedModels}, generating bitstrings as a quantum-enhanced Markov Chain Monte Carlo sampler \cite{Layden2023Quantum-enhancedCarlo} (MCMC). More specifically, the quantum sampler generates new bitstrings which are decoded into designs (Figure \ref{fig:latent_MCMC}). During the decoding step, the cost function of the designs is energy-matched to the energy function of the bitstrings using an adversarial discriminator loss (Figure \ref{fig:latent_MCMC}.a). If the energy matching loss is low, an MCMC scheme will efficiently sample a Boltzmann distribution in the native space and in turn the non-native design space, as weighted by their matched cost functions.

Energy matching on a non-native problem is demonstrated by training an AQAM to sample high-efficiency designs for periodic metasurfaces that provide a broad-angle filtering function \cite{Yu2014FlatMetasurfaces}. We test the AQAM's performance through an ablation study by swapping its quantum sampler with classical samplers, namely a single bit-flip sampler and a uniform sampler, and demonstrate near-term comparative advantage by measuring the Renyi divergence of an AQAM's quantum-enhanced encoder \cite{vanErven2012RenyiDivergence} against an ideal Boltzmann distribution. We also demonstrate large temperature regions where the AQAM has a larger MCMC spectral gap compared to its classical counterparts, thereby demonstrating both an asymptotic and near-term performance benefit over similar classical samplers.

\section{Autoencoding Non-Native Distributions}

Consider a dataset of designs $\{\x_l\} \subset {X}$ sampled i.i.d. from the data distribution $p(\x)$ over the design space $X$. Additionally, the design space includes an objective function $C_\x: X \rightarrow \mathbb{R}$, so that the sample probability is a Boltzmann distribution

\begin{align}
  p(\x) = \mu(\x; C_\x, \tau) \triangleq \frac{e^{-C_\x(\x)/\tau}}{Z_{X}}
\end{align} 

for some temperature $\tau$ and partition function $Z_{X}=\sum_{\x} \exp(-C_\x(\x) / \tau)$. Additionally, consider a ``native" latent space $\{0,1\}^n$, which represents the measurement outcomes from a qubit-based quantum computer, e.g. bitstrings $\z \in \{0,1\}^n$. We assume a priori that these bitstrings have some classical energy cost $C_\z(\z)$ and are sampled from a Boltzmann prior $q(\z)=\mu(\z;C_\z,\tau)$.

The goal of the AQAM is to find a learned conditional decoder distribution $p_\theta(\x \,\vert \,z)$ that maximizes the log-likelihood of our model $p_\theta(\x)$, i.e., $\mathbb{E}_{\z \sim q(\z)}[\log p_\theta(\x\vert z)]$, of generating designs under the Boltzmann distribution. Maximizing the log-likelihood is notoriously difficult due to the intractable problem of computing the aggregate posterior $\sum_{\x \in X}q(\z \vert \x)$ \cite{Kingma2013Auto-EncodingBayes, Khoshaman2019GumBolt:Priors}.
Instead, the AQAM uses an autoencoder approach by bounding the log-likelihood with the evidence lower bound and introducing an encoder distribution $q_\ph(\z \vert \x)$ to approximate the Boltzmann prior $q(\z)$ under the Kullbach-Liebler (KL) divergence \cite{Kingma2013Auto-EncodingBayes,Makhzani2015AdversarialAutoencoders}.

However, directly training a quantum-enhanced encoder distribution to approximate the prior with variational methods introduces barren plateaus and noisy sampling problems \cite{Khoshaman2018QuantumAutoencoder, Wang2021Noise-inducedAlgorithms,Cerezo2021CostCircuits, Cerezo2021VariationalAlgorithms}. 
Therefore, the AQAM uses an adversarial loss function given by  
\begin{align}
\log p_\th(\x)\geq \argmax_{\ph, \th}\mathbb{E}_{\z \sim q_\ph(\z \vert \x)}\Big[\log p_\th(\x \vert \z)& \label{eq:aae} \\ - \sum_{\x' \sim p_\th(\x' \vert \z)}\vert C_\z(\z) -  C_\x(\x')\big\vert ^2\Big]& \nonumber.
\end{align} 
The first term is the reconstruction loss, which is the likelihood that a sample $\x$ fed through the encoder will be decoded into the original sample $\x$.
The second term is the discriminator loss $D(\z, \x)$ and regularizes the non-native energy distribution to be energy-matched to the native energy distribution. 
This way, realizing a Boltzmann sampler through the prior will enforce a Boltzmann distribution in the non-native space.

In the original formulation of the adversarial autoencoder, the adversarial posterior is realized using a discriminator network that classifies each native sample $\z$ as being either distributed according to the desired prior $q(\z)$ or the encoder \cite{Makhzani2015AdversarialAutoencoders}.
Instead of using a discriminator network, the AQAM guarantees the encoder converges to a Boltzmann distribution by a MCMC adversarial channel sampling step $\Gamma(\z\vert \z')$ which obeys detailed balance. Then, the adversarial posterior distribution is given by
\begin{equation}
    q_\phi(\z\vert \x) = \sum_{\z'} \Gamma(\z\vert \z')\tilde q_\phi(\z'\vert \x), \label{eq:adv_post}
\end{equation}
where $\tilde q_\phi(\z'\vert \x)$ is a classical encoder and the channel $\Gamma$ uses a quantum sampler.
If the channel $\Gamma$ obeys detailed balance using the Metropolis-Hastings update rule, then an iterative Markov chain Monte Carlo algorithm will eventually converge to a Boltzmann distribution with a mixing time inversely proportional to the spectral gap \cite{Levin2006MarkovTimes}.
We take inspiration from the reparameterization approach and Markov chain diffusion process in the original variational autoencoder (VAE) \cite{Kingma2013Auto-EncodingBayes} and diffusion models \cite{Ho2020DenoisingModels}, respectively, to construct $\Gamma$. 
Firstly, the AQAM's quantum encoder $q_\ph$ is similar in spirit to the original VAE's encoder that realizes a Gaussian prior by using the reparameterization trick \cite{Kingma2013Auto-EncodingBayes} to avoid difficult gradient calculations. The classical encoder $\tilde{q}(\z' \vert \x)$ samples the initial point $\z'$, which is analogous to the original VAE's encoder sampling the Gaussian mean $\vec x$.
The AQAM's encoder breaks the dependence on the internal distribution of $\Gamma$ by a change-of-variables so that the sampled native state $\z$ acts as additive ``noise" from the adversarial channel and is independent of the output during training
\begin{equation}
    \epsilon = z' \oplus z \label{eq:add_noise}
\end{equation}
where addition is defined modulo 2, with a correspondingly redefined channel $\Gamma(\epsilon \vert  z')$.
This step is analogous to the VAE's encoder sampling from the independent Gaussian with mean $0$ and covariance $\sigma/\tau$ which is then shifted by the encoder's computed mean $\vec x$.
However, unlike the VAE's encoder, we take the channel $\Gamma(\z\vert \z')$ to be a complete ``black box" with only the knowledge that $\Gamma$ satisfies detailed balance. 
 In this way, the total loss can be written as a nested average with respect to the noise
\begin{multline}
    \mathbb{E}_{\z'\sim \tilde{q}_\phi(\z'\vert \x)}\mathbb{E}_{\epsilon\sim \Gamma(\epsilon \vert  \z')}\Big[\log\big[p_\theta(\x \vert  \z' \oplus\epsilon )\big]\\- \mathbb{E}_{\x'\sim p_\theta(x'\vert \z'\oplus \epsilon)}\big\vert C_z(\z' \oplus \epsilon) - C_\x (\x') \big\vert ^2
    \Big]. \label{eq:noisy_recon}
\end{multline}

Secondly, we take inspiration from the Markov chain structure in diffusion models.
If we extend a Markov chain of updates to the encoder's state $\z$ such that the (Gaussian-variant) channel satisfies the detailed balance condition using an energy function $E(\z)=\sigma \z^2$, then the encoder's underlying distribution will be a Gaussian distribution with mean $0$ and variance $\sigma$, minimizing the KL-divergence to the desired prior Gaussian.
As the variance of the noise $\epsilon$ is increased, the mutual information of the channel $\Gamma(\z\vert \z')$ decreases and the reconstruction loss increases.
Alternatively, the amount of volume locally explored by the native bit strings $\z' \oplus \epsilon$ is increased, generating novel non-native designs to better train the energy-matching term.


\section{Training}

Training an AQAM requires a few extra steps to enable the typical autodifferentiable gradient-based approaches of classical machine learning \cite{Goodfellow2016DeepLearning}. First, the discrete bitstrings $z'\in\{0, 1\}^n$ are relaxed into continuous and differentiable real numbers $z'\in \mathbb{R}$ using the hyperbolic tangent function, a change-of-basis, and a quantizer to clamp the output value between 0 and 1 \cite{Rolfe2017DiscreteAutoencoders, Fajtl2020LatentAutoencoder} and ensure backpropagation through the discrete bitstring variables generated by the channel.

Second, the action of the channel $\Gamma(\epsilon|z')$ is applied simply by sampling $\epsilon\sim \Gamma(\z \vert z')$ and then ``detaching" the computation graph of $\z$ to generate $\epsilon$, much like the noise introduced in the reparameterization trick of variational autoencoders \cite{Kingma2013Auto-EncodingBayes}. In order to compute gradients while avoiding barren plateaus, this step implicitly assumes $\partial_{z'}\Gamma(\epsilon \vert z')=0$.

Finally, training of the variational parameters $\theta$ and $\phi$ is done by backpropagation-enabled gradient descent of equation \ref{eq:noisy_recon}.
Each step samples from the training space of designs $\{\x\}$, which are each encoded into a single maximum-likelihood sample $\z'$ from $\tilde{q}_\phi(\z'\vert \x)$ \cite{Kingma2013Auto-EncodingBayes,Rolfe2017DiscreteAutoencoders} (Figure \ref{fig:latent_MCMC}.b).
Then, the channel $\Gamma(\epsilon|z')$ generates a small set of noise $\epsilon$ per sample $\z'$. Due to detaching the noise $\epsilon$ from the computation graph, equation \ref{eq:noisy_recon} is autodifferentiable, and the parameters $\theta$ and $\phi$ are updated using gradient descent. See appendix \ref{app:training} for more details.


\section{Update Samplers}

A key part of the AQAM is the update sampling step $\Gamma(\epsilon|z')$, which ultimately proposes a new design $\x$ by scrambling the encoded data in the latent space with adversarial noise $\epsilon$. To ensure Boltzmann convergence, the update sampling step should obey detailed balance $\Gamma(z|z') = \exp\big((C_z(z) - C_z(z'))/\tau\big)\Gamma(z'|z)$ with respect to the native energy function $C_z$ at temperature $\tau$ for all $z$, $z'$.
By obeying detailed balance, repeated sampling using Markov chain updates 
\begin{align}
  \Gamma_f(\z \vert \z') \triangleq \sum_{\z_{f-1}}\Gamma(\z \vert \z_{f-1})...\sum_{\z_{1}}\Gamma(\z_1\vert \z'). \label{eq:chainR}
\end{align}
will eventually converge to a stationary Boltzmann distribution \cite{Levin2006MarkovTimes}. 
By aligning the adversarial noise $\epsilon$ with an MCMC step that samples from a Boltzmann distribution, as well as enforcing an energy correlation between non-native space and design samples through the discriminator loss term, the distribution of samples in the design space will also be Boltzmann distributed.
One metric often used for benchmarking a MCMC proposal sampler is the spectral gap, i.e., the difference between the two largest eigenvalues of the transfer matrix $P_{z, z'} = \Gamma(z|z')$ \cite{Levin2006MarkovTimes}, which is inversely proportional to the number of MCMC steps required to mix and generate an unbiased sample from the Boltzmann distribution. A larger spectral gap leads to faster mixing and thus a more efficient sampling of novel designs. 

Depending on the desired mode of the AQAM, the channel $\Gamma(z|z')$ may be modified as part of the training process to bias toward higher quality designs or a larger variance of designs.
For example, during initial training, the temperature $\tau$ and depth $f$ can be fixed so that the mutual information between the input and output of the channel can initially be large (or optimal, given an update rule $\Gamma(z|z')=\delta_{z,z'}$).
As we increase the training time, we can lower the MCMC temperature, thereby lowering the effective search space of samples being learned by biasing to higher-quality designs.
Then, by repeated resampling of the low-temperature region, we generate optimized non-native designs.
Alternatively, if we desire to explore the non-native space, we can instead increase the temperature and channel depth to increase the generative variance. 
By increasing the channel depth slowly, we also increase the training variance and Boltzmann convergence. We realize this increased depth through ``telescoping'' the update sampling step $\Gamma_f$ by adding multiple atomic MCMC update steps $\Gamma$ into a composite step

\begin{equation}
    \Gamma_{f+1}(z|z') = \sum_{z''}\Gamma(z|z'')\Gamma_{f}(z''|z').
\end{equation}

Such a ``telescoping'' step reduces the mutual information of the channel and the reconstruction quality. However, reducing the mutual information increases the number of novel samples and thus novel designs, which lets the autoencoder reduce the energy matching discriminator loss. By resampling the same stationary point $\z'$,  designs with similar figures of merit to $\z'$ can be explored.

We realize each atomic sampling step using the Metropolis-Hastings update rule \cite{Levin2006MarkovTimes}

\begin{align}
    \Gamma(\z\vert \z') = \nonumber \\ r(\z\vert \z')\times \text{MIN}\big[1,\exp(-(C_\z(\z) - C_z(z'))/\tau ) \big], \label{eq:acceptance_rule}
\end{align}
where $r(z|z')=r(z'|z)$ is a symmetric proposal sampler. 
There are many choices for proposal samplers. We consider a quantum sampler, a classical single bit-flip sampler, and a uniform sampler, denoted by $r_q$, $r_l$, and $r_u$, respectively. The proposal samplers are the main source of noise $\epsilon$ and directly control the quality of MCMC sampling.

\subsection{Quantum proposal sampler}\label{sec:quantum_sampler}

\begin{figure}
  \begin{center}
    \includegraphics[scale=0.8]{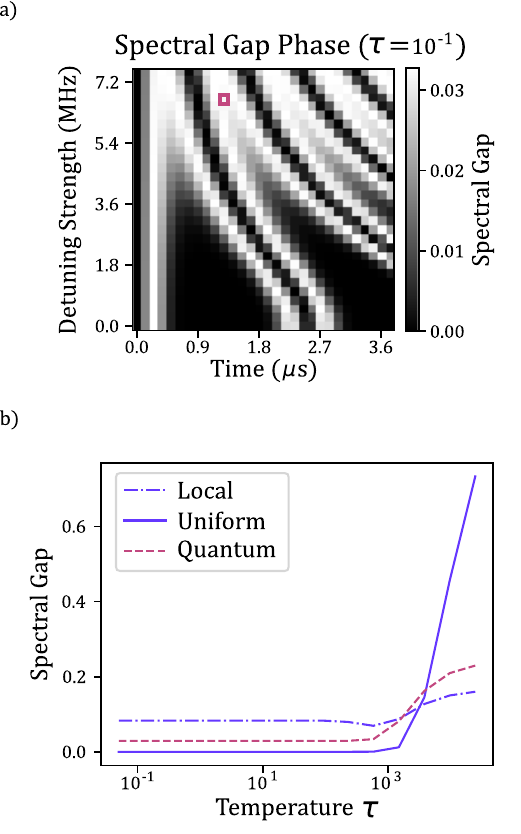}
  \end{center}
  \caption{\textbf{Spectral properties of classical and quantum MCMC samplers} (A) A phase plot shows the spectral gap of a neutral atom quantum sampler with fixed inverse temperature $\tau = 10^{-1}$ for varying detuning strengths $\Delta \in  [0, 7.8)$ (MHz) and evolution times $t_q \in [0, 3.8) \mu$s. The selected parameters for the quantum channel sampler are enclosed in the purple square. (B) Temperature $\tau \in [10^{-1}, 10^5]$ compares the spectral gap $\delta$ of the quantum $r_q$, single bit-flip $r_l$, and uniform $r_u$ proposal samplers.
  }
  \label{fig:phase_plot}
\end{figure}
The AQAM's quantum proposal sampler $r_q$ is based on quantum evolution from the current basis product state $\vert z'\rangle$, which encodes the bitstring $z'$, under some unitary $U$ such that
\begin{equation}
    r_q(z \;\vert z') \;=\;\big\vert \langle z \vert  U \vert  z'\rangle \big\vert ^2, \label{eq:q_update}
\end{equation}
where the probability of selecting a new bitstring $z$ is given by measuring the quantum state $U\vert z'\rangle$ in the z-basis \cite{Layden2023Quantum-enhancedCarlo, Orfi2024BoundingCarlo}. 
In this work, we use a unitary based on a time-independent quench of an analog mode neutral-atom quantum computer \cite{Ebadi2021QuantumSimulator} with the generator given by the Rydberg atom Hamiltonian
\begin{multline}
    U = \exp\big(-it(\\
    \sum_i \frac{\Omega}{2}(\vert 1\rangle \langle 0\vert  +\vert 0\rangle\langle 1\vert)  - \Delta \hat n_i + \sum_{ij} V_{ij} \hat n_i \hat n_j)\big), \label{eq:Qunitary}
\end{multline}
\\
where the Van der Waals interaction $V_{ij}=C_6 \vert \vec r_i - \vec r_j\vert ^{-6}$ is based on the relative positions $\vec r_i, \vec r_j$ of each Rb$_{87}$ atom $i, j$ as the qubit hosts, and $\hat n_i = \vert 1_i\rangle\langle 1_i\vert $.
For more details, see \cite{Wurtz2023Aquila:Computer}.
The state $\vert z\rangle$ is prepared using an adiabatic protocol where atoms in the $0$ state are masked by a strong local detuning field generated by a Stark shift from an individually targeted 795nm laser.

While contemporary hardware can implement dynamics on up to 256 qubits, we select spatial geometries of King's subgraphs of 12 qubits simulated using exact diagonalization with the Bloqade package \cite{Bloqade.jl:Architecture.}.
As shown in Figure \ref{fig:phase_plot}, we select the optimal quantum sampler (\ref{eq:q_update}) parameters $\Delta = 1.06 \cdot 2 \pi \text{MHz}$ and quench time $t_q = 1.29 \mu\text{s}$  by performing a hyperparameter sweep of the spectral gap $\delta$ and then choosing a point with both a large spectral gap and a large margin of error as indicated by the purple square in Figure \ref{fig:phase_plot}.a.   

The classical energy $E(\z)$ of the state $z\in \{0,1\}^n$ is computed as
\begin{align}
  E(\z) = \sum_i\Delta z_i + \sum_{i,j} V_{ij}z_iz_j,\label{eq:class_en}
\end{align}

For $\Omega t=0$, there are no dynamics, and the proposal probability $r_q(z\vert z')=\delta_{z, z'}$. For $\Omega t\to 0$ perturbatively, there is a small chance of a single bit-flip, so the proposal samples from bitstrings are at a Hamming distance 0 and 1. For $\Omega t\to\infty$ the dynamics are typically ergodic, preserving only the total energy, which means that each term in the Hamiltonian will have a finite variance. If the Rabi term $\Omega$  in (\ref{eq:Qunitary}) is much smaller than the classical energy term in (\ref{eq:class_en}) (as is the case for strong neighboring interactions $V_{ij}$), the final classical energy will be close to the initial energy, i.e., $E(z)\approx E(z')$. This is a key advantage of using a quantum quench as a variational update rule: for $\Omega t\to \infty$ the new bitstring will be completely unbiased from the initial, it will nonetheless have a similar classical energy, ``tunneling" through energy barriers to uniformly sample low-energy samples, thereby enforcing the energy matching condition much more easily. In this way, varying the evolution time of a quench is the same as ``telescoping" the MCMC step, except with the possibility of a smaller effective spectral gap \cite{Levin2006MarkovTimes, Layden2023Quantum-enhancedCarlo}.

\subsection{Classical proposal sampler}

 We measure the AQAM's performance through an ablation study with two classical-only proposal samplers. The first is a single bit-flip sampler \begin{align}r_l(\z \vert \z') =  \delta(H(\z, \z') - 1) / n \label{eq:l_update},\end{align} 
where $H(\z, \z')$ is the Hamming distance from $\z$ and $\z'$. The second is a uniform sampler 
\begin{align}r_u(\z \vert \z') = 2^{-n}.\label{eq:u_update}\end{align}

The single bit-flip proposal sampler can be considered the $\Omega t\to 0$ ``no quantum" limit of the quench sampler of Sec.~\ref{sec:quantum_sampler}. In this way, the classical and quantum performance can be directly compared to demonstrate a ``comparative advantage" \cite{Wurtz2024SolvingAlgorithms}.

\begin{figure}[t!]
  \begin{center}
    \includegraphics[scale=0.7]{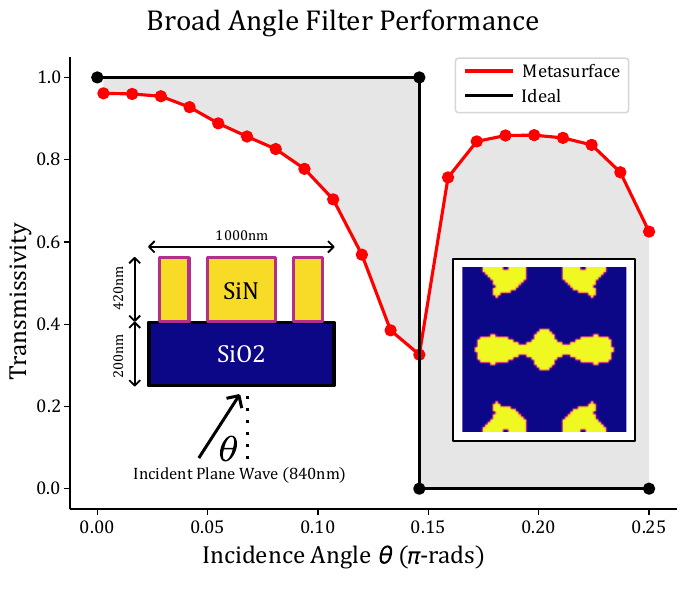}
  \end{center}
  \caption{\textbf{Broad Angle Filter}: a periodic dielectric metasurface with a $\text{SiO}_2$ substrate and $\text{SiN}$ nanopillars. The filter reflects $840$nm light at an angle of incidence $\theta > 0.14$.
  The substrate and nanopillars thicknesses are 200nm and 420nm, respectively, with a unit cell width / lattice spacing of 1000nm.
  Transmissivity is plotted for the unit cell $T_u(\theta)$ (red line) and the ideal filter $T_i(\theta)$ (black line). The non-native objective function is the difference between the two lines as shaded in grey.}
  \label{fig:aoi_filter}
\end{figure}

\begin{figure*}[t!]
  \begin{center}
    \includegraphics[width=\textwidth]{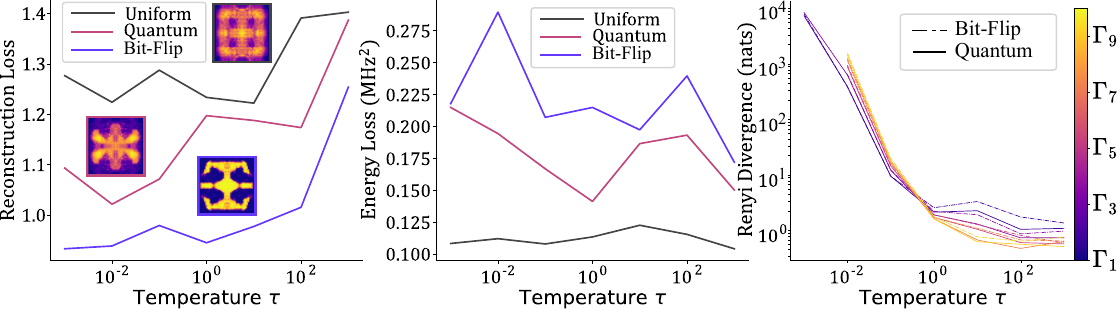}
  \end{center}
  \caption{
    \textbf{AQAM Performance for Boltzmann Sampling of Broad Angle Filters:} each point in plots (A) and (B) is averaged over the final 200 training epochs of 10 models with randomly initialized parameters $\th$ and $\ph$ (Figure \ref{fig:encoder_decoder}), a constant training
    temperature $\tau \in [10^{-3}, 10^{3}]$, and a constant channel depth of $3$, i.e., $\Gamma_3$.
    (A) The binary cross-entropy reconstruction loss (\ref{eq:aae}) is measured for increasing training temperatures.
    The bit-flip sampler's lower reconstruction loss indicates the designs and bitstrings have more mutual information. Above each sampler's plot is an example design representative of the reconstruction quality for that sampler.
    (B) The discriminator loss $D(\z, \x)$ is plotted for the same models as in (A). The quantum sampler has lower energy loss, indicating better energy matching between the design space cost function $C_\x$ and native energy function $C_\z$.  
    (C) The Renyi divergences (\ref{eq:renyi}) for varying channel depths $\Gamma_1, ..., \Gamma_9$ (color gradient) are plotted against the validation temperatures $\tau$ for both the single bit-flip (dash-dot line) and quantum (solid line)  proposal samplers. The smaller Renyi divergence for the quantum proposal sampler and increasing channel depth indicates stronger convergence to a Boltzmann distribution over the validation samples.
  }
  \label{fig:results}
\end{figure*}

\section{Meta-Optical Device Design}
Dielectric metasurfaces are low-loss, ultra-thin, flat-optical structures comprising subwavelength ``meta-atom'' unit cells, which can provide arbitrary control over the phase, amplitude, polarization, and optical impedance of a wavefront \cite{Wilson2021MachineProblems, Hsu2022Single-AtomTweezer, Overvig2019DielectricPhase, Yu2014FlatMetasurfaces} (Figure \ref{fig:aoi_filter}).  
Advanced unit cell topology optimization techniques like autoencoder-based latent optimization \cite{Kudyshev2020RapidLearning} and quantum annealing \cite{Wilson2021MachineProblems} have greatly improved the computation time and overall design efficiency of photonic metasurfaces as compared to traditional gradient descent and adjoint optimization techniques (546x speed-up with 1.05x improvement in efficiency \cite{Wilson2021MachineProblems}). 
As a showcase example, we demonstrate the AQAM by optimizing the unit cell topologies of a broad-angle filter, which is designed to reflect light when the angle of incidence is above a given threshold angle $\theta$, thereby providing smaller beam waists and less power drift \cite{Ebadi2021QuantumSimulator}.
The broad-angle filter uses a $\text{SiO}_2$ substrate and SiN nanopillars, which are non-absorptive at the operation wavelength of 840nm, and shows a high-efficiency broad-angle filtering functionality for angles $\theta > 0.14\pi$. 
The figure-of-merit for the studied metasurface is given by the integral
\begin{align}
    C_\x(\x) = 1 - \int_{\theta}(T_i(\theta) - T_u(\theta, \x))d\theta \label{eq:cx}
\end{align}
where $T_i(\theta)$ is the ideal transmissivity and $T_u(\theta, \x)$ is the computed transmissivity for the unit cell $\x$ at the angle of incidence $\theta$.

\section{Numerical Simulations}

We demonstrate the AQAM with an ablation study (Figure \ref{fig:results}) using a benchmarking process outlined in appendix \ref{app:benchmark}.
The quantum proposal sampler is modeled numerically by simulating a 12-atom Rydberg system using the Bloqade emulation package \cite{Bloqade.jl:Architecture.} to directly compute the transition matrix $r=|U|^2$.
The same architecture (Figure \ref{fig:encoder_decoder}) is also compared with the classical samplers with new, randomized initial weights $\theta$ and $\phi$ as a no-quantum limit \cite{Wurtz2024SolvingAlgorithms}.
Results are shown in Figure~\ref{fig:results}. For each proposal sampler, we generate a maximum likelihood initial sample $\z_0 \sim q_\ph(\z_0 \vert \x)$ and then sample the channel $\Gamma_3(\z_f \vert \z_0)$ to generate several samples with native energy close to $\z_0$.  
As indicated by both the reconstruction loss and discriminator loss (Fig.~\ref{fig:results}.a and Fig.~\ref{fig:results}.b), the single bit-flip sampler (\ref{eq:l_update}) preserves the most mutual information between the designs $\x$ and the native bitstrings $\z_0 ... \z_f$ while not preserving the energy matching condition as well as the other samplers.
However, the uniform sampler (\ref{eq:u_update}) loses nearly all \footnote{rarely the encoder chooses a low energy native sample $\z_0$ which does not update during MCMC} mutual information due to the statistical independence of the proposed sample $\z_{t+1}$ from the current bitstring $\z_t$ while maintaining the best energy matching due to the large variance in bitstring samples and the ``averaged design'' generated by many uniformly sampled bitstrings (Figure \ref{fig:results}.a). 

The main comparative advantage is how well the AQAM samples a Boltzmann distribution in the non-native space $X$, as measured by the empirical Renyi divergence (Figure \ref{fig:results}.c) and the spectral gap (Figure \ref{fig:phase_plot}.a.) The Renyi divergence (\ref{eq:renyi}) measures the divergence, in natural units (nats), between the measured non-native sample distribution compared to an ideal distribution $\mu(\x; C_\x, \tau)$ \footnote{Meaning a lower Renyi divergence means stronger convergence to the ideal Boltzmann distribution}. 
We assume the target distribution is a Boltzmann distribution over the generated designs in the non-native space. 
The AQAM's decoder space is localized to a subset of designs $\hat{X}$ which it can generate given an input bitstring $\z \in \{0, 1\}^{12}$.
For validation purposes, it is intractable to measure the output distribution of the decoder $\hat{X}$ relative to a true Boltzmann distribution over the entire design space $X$. 
Instead, we measure the relative distribution of the AQAM to generate a Boltzmann distribution over $\hat{X}$ given a set of native bitstring samples, as indicated by the Renyi divergence in Figure \ref{fig:results}.c. 

This means we can use this benchmark for measuring comparative advantage in realizing a Boltzmann distribution over the generated design space relative to a decoder's output. During training, we apply the Metropolis-Hastings acceptance rule over the native samples $\z$. However, for validation we apply the Metropolis-Hastings acceptance rule to the objective function values $C_\x(\x)$ for each proposed design $\x$, see Appendix \ref{app:benchmark}. As indicated in Figure \ref{fig:results}.c, the quantum sampler achieves a lower Renyi divergence as compared to the classical bit-flip sampler, and therefore achieves stronger convergence to a Boltzmann distribution in the non-native space. Additionally, we see that by increasing the channel depth, the Renyi divergence decreases for both samplers.
To improve the performance on this benchmark, the native channel $\Gamma_3$ needs to be energy-matched to the non-native design space so that samples in the native space correlate to the non-native space.
As indicated by Figure \ref{fig:results}.b, the quantum sampler achieves stronger energy-matching as compared to the single bit-flip sampler. This is corroborated by the larger spectral gap for larger temperatures $\tau$, indicating strong asymptotic convergence to a Boltzmann distribution \cite{Layden2023Quantum-enhancedCarlo}. Despite the uniform sampler's larger spectral gap in the same large $\tau$ region, the loss of mutual information makes it the worst update rule for autoencoding.

While the reconstruction loss demonstrated (Figure \ref{fig:results}.a) shows a lower reconstruction quality for the quantum sampler compared to a single bit-flip sampler, this is likely due to the small number of native sample qubits.
As we increase the number of qubits, the reconstruction quality is expected to approach the single bit-flip sampler quality.
However, we expect this is not the case for the uniform sampler because the mutual information is lost due to the sampler being statistically independent from the current sample.
Because sampling quantum simulations can be computationally expensive, we only considered a fixed native energy model $C_\z(\z)$.
However, one could always variationally train the classical energy model in place of a parameterized quantum circuit \cite{Benedetti2019ParameterizedModels, Wilson2021MachineProblems} to increase the mutual information.
Additionally, by using tensor network sampling \cite{Rudolph2023SynergisticNetworks, Hibat-Allah2021VariationalAnnealing} or single bit-flip samplers, an AQAM can be pretrained before using higher-quality quantum samples.
Once an AQAM is pretrained on a classical sampler, the weights of the encoder and decoder can be transferred to a full quantum sampler.

Notably, generating a Boltzmann distribution over designs only occurs if a) the energy discriminator term is zero, and b) the decoder uniformly maps equal-energy samples to designs. If the decoder is poorly trained on a subset of the design space or the energy discriminator term is nonzero, the output distribution may be far from the Boltzmann target.
Likewise, we considered only 50 designs for training due to the empirical training quality influenced by the variance of latent samples and small qubit size under various temperatures and channel depths.
We expect that increasing the number of qubits under the same channel depth and temperatures will increase the reconstruction quality for a larger training set size while outperforming the classical models on Boltzmann convergence.

\section{Conclusion}

One of the main challenges facing quantum generative modeling is finding efficient ways of using bitstrings generated by existing quantum samplers for sampling modern applications which don't {\it natively} use bitstrings, such as natural language processing, generative art, and engineering inverse design.

This work demonstrates a novel quantum generative machine learning model known as the adversarial quantum autoencoder model (AQAM) that uses the native bitstrings of quantum samplers in the latent space of an adversarial autoencoder model for sampling Boltzmann distributions over engineering design problems. 
The proposed AQAM framework samples the latent space of a discrete adversarial autoencoder using quantum-enhanced MCMC while ensuring that the energy of the latent samples matches, up to constant factors, the optimization landscape's cost function.
We demonstrate the AQAM on 12 qubits, which enables exact computation of the spectral gap and Renyi divergence to demonstrate comparative advantage against comparable classical models.
The next steps would be to determine how to choose the ideal number of qubits for a given design space, either through sparsity-aware models or mutual information analysis. 
A common problem facing generative modeling tasks is the choice of the modeling distribution, which can only be assessed by analysis of the log-likelihood function or other convergence metric.
The advantage of using MCMC in the latent space is that, generically, any arbitrary distribution can be realized.   
  
We note that while AQAM was demonstrated using quantum resources, this approach can be applied to both classical and quantum samplers. 
By using a classical single bit-flip sampler, the size of latent bitstrings can be increased arbitrarily, alleviating the low reconstruction quality and training set size. 
However, as we demonstrated on a small, 12-qubit system, there may be stronger Boltzmann convergence when using a quantum sampler. This would be a topic of future research.  
By being a hybrid quantum-classical model, the AQAM opens the door for comparative advantage in non-native quantum-enhanced generative modeling, allowing for near-term quantum and classical samplers to be incorporated into the modern generative AI pipeline.

\section{Acknowledgements}

This work was supported by the DARPA ONISQ program (Grant No. W911NF2010021) and DARPA-STTR award (Award No. 140D0422C0035), the National Science Foundation award 2029553-ECCS, DARPA/DSO Extreme Optics and Imaging (EXTREME) Program (HR00111720032), and the U.S. Department of Energy (DOE), Office of Science through the Quantum Science Center (QSC), a National Quantum Information Science Research Center.
The authors thank Alexander Keesling and Phillip Weinberg for helpful discussions.

\bibliographystyle{apsrev4-1}
\bibliography{references}

\appendix

\section{Training}
\label{app:training}
\begin{figure}
  \begin{center}
    \includegraphics[scale=1.0]{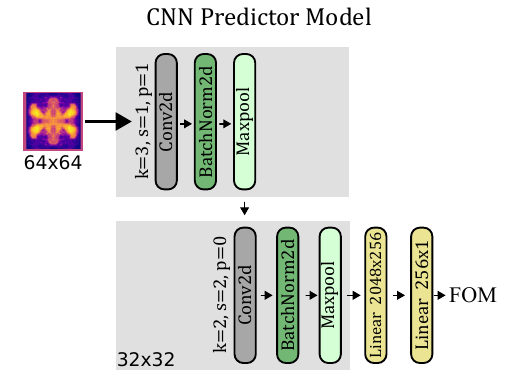}
  \end{center}
  \caption{{\bf CNN Predictor Model:} is a convolutional neural network (CNN) model trained to compute $C_\x$. The model utilizes 2D convolutions (gray), 2D batch normalization layers (dark green), Gaussian Error Linear Unit (GELU) activations, and maxpool layers (light green). All layer parameters are the same in each layer, where applicable, with the parameter labels being $k$-kernel size, $s$-stride, and $p$-padding.}
  \label{fig:cnnattention}
\end{figure}

The training dataset designs are transferred from topology optimized unit cells for a thermophotovoltaic application \cite{Wilson2021MachineProblems}. 
We first compute the non-native objective function $C_\x$ using (\ref{eq:cx}) for 10,000 images of 64px $\times$ 64px unit cells using Rigorous Coupled-Wave Analysis (RCWA) calculations and then train a convolutional neural network predictor \cite{Simonyan2015VeryRecognition} as a faster solver with $0.01$ mean-squared-error loss against the simulations (Figure \ref{fig:cnnattention}). 

Then, we randomly select $50$ images \footnote{This was empirically chosen and is likely small due to the small native state size.} for training. We randomly initialize the parameters $\th$ and $\ph$ for the classical encoder $\tilde q_\ph(\z \vert \x)$  and decoder $p_\th(\x \vert \z)$, which are implemented using attention-based convolutional neural networks (Figure \ref{fig:encoder_decoder}). 
For each temperature $\tau \in \{10^{-3}, 10^{-2}, ..., 10^3\}$, we consider a prior $\mu(\z; C_\z, \tau)$ where the native energy function $C_\z$ is given by the classical energy (\ref{eq:class_en}) of the chosen neutral atom system for the hyperparameters $\Delta = 1.06 \cdot 2 \pi \text{MHz}$ and quench time $t_q = 1.29 \mu\text{s}$ as outlined in Figure \ref{fig:phase_plot}. 

We take each design sample $\x$ and encode $\z_0 \sim \tilde q_\ph(\z_0 \vert \x)$ an initial native sample bitstring $\z_0$ in the spin basis $\z_0 \in \{-1, 1\}^n$ \footnote{to avoid any $\mathbf{0}$ vector bias} using an Ising change of basis of the energy function $C_\z$, see \cite{Lucas2014IsingProblems, Wilson2021MachineProblems}.
We relax the discrete bitstring $\z_0$ into a continuous bitstring on the backward pass using a hyperbolic tangent function and a quantizer \cite{Fajtl2020LatentAutoencoder}.
This ensures we can backpropagate through the discrete bitstring variables $\z_0$.
To apply the channel $\Gamma(\epsilon \vert \z_t)$, the bitstring $\z_t$ is converted to the $\{0, 1\}$ basis. 
Then, we sample the maximum likelihood noise vector $\epsilon_t \sim \Gamma(\epsilon_t \vert \z_t)$ and apply the output $\epsilon_t$ to $\z_t$ by assuming $\partial_{z_t} \Gamma(\epsilon_t \vert  \z_t)=0$. We do so by ``detaching'' the computation graph of $\epsilon_t$, much like the noise introduced in the reparameterization trick \cite{Kingma2013Auto-EncodingBayes}.
Finally, we generate the original design $\x$ using the decoder $p_\th(\x_t \vert \z_t)$, compute the loss functions in (\ref{eq:noisy_recon}), and optimize the parameters $\th$ and $\ph$ using gradient descent and backpropagation.

\section{Benchmarks}
\label{app:benchmark}
For the reconstruction loss and energy loss benchmarks in Figure \ref{fig:results}, we choose a different update sampler and sweep the inverse temperature $\tau$ from $\tau = 10^{-3} \rightarrow 10^4$. For each value of $\tau$, we train 10 models, each initialized with different random parameters, and average the benchmark score for the last 200 epochs of each model.
\subsection{Renyi Divergence}
\label{app:renyi}
We benchmark the proposed AQAM's empirical convergence to a Boltzmann distribution using the Renyi divergence \cite{vanErven2012RenyiDivergence}.
We first compute an ideal Boltzmann distribution over all sampled designs from the decoder by computing the partition function $Z_X = \sum_{\z \in \{0, 1\}^{12}}\sum_{\x \sim p_\th(\x \vert \z)}\exp(-C_\x(\x)/\tau)$.
Next, we bin the cost functions into regions $\{(C_0, C_1), ..., (C_i, C_{i+1}), ...\}$.
For each region $i$, we compute the target probability $\hat{\mu}_i(\x)$ of a sample $\x$ being in a given cost function region $(C_i, C_{i+1})$, i.e., we compute $ \hat{\mu}_i(\x) = P(C_\x(\x) \in (C_i, C_{i+1})) = \sum_{\x : C_\x(\x) \in (C_i, C_{i+1})} \frac{\exp(-C_\x(\x)/\tau)}{Z_X}$.   
Then, we sample the empirical probability $\hat{p}_i(\x)$ of each generated sample $\x \sim p_\th(\x \vert \z_f)\Gamma_3(\z_f \vert \z_0)q_\ph(\z_0)$ where $q_\ph(\z_0)$ is marginalized over the encoder conditioned on the training dataset. To compute  $\hat{p}_i(\x)$, we take 5000 samples at the specified channel depth and temperature $\tau$.
Then, given the sample distribution $\hat{p}_i(\x)$ against an ideal Boltzmann distribution $\hat{\mu}_i(\x)$, we compute the Renyi divergence as
\begin{align}
    D_\alpha(\hat{p} \vert\vert \hat{\mu}) = \frac{1}{1-\alpha}\ln{\sum_{i = 1}^{n}\hat{p}_i(\x)^\alpha\hat{\mu}_i(\x)^{1-\alpha}} \label{eq:renyi}
\end{align}
where the order-parameter $\alpha \neq 1$. The Renyi divergence is an empirical approximation to the KL-Divergence such that 
\begin{align}
    \lim_{\alpha \rightarrow 1} D_\alpha(\hat{p} \vert\vert \hat{\mu}) = D_{\text{KL}}(\hat{p} \vert\vert \hat{\mu}).
\end{align}
Because the Renyi divergence is discontinuous at $\alpha = 1$, we use an order parameter $\alpha = 0.999$. 
For validating the Renyi divergence, we sample the native samples $\z \sim \mu(\z; C_\z, \tau)$ using both the quantum and bit-flip proposal sampler on each sampler's pretrained models, i.e., each model trained with the bit-flip proposal sampler was validated using both a quantum and a bit-flip proposal sampler and vice versa. The reported score is the average Renyi divergence for both models trained with a bit-flip and a quantum proposal sampler. Thereby showing that the quantum sampler performs better than the bit-flip sampler on models trained with the bit-flip sampler. 

\subsubsection{Validation Acceptance Rule}
For the Renyi divergence benchmark, we sample $\hat{p}$ by using the Metropolis-Hastings acceptance rule over the design space, i.e.,
\begin{align}
    \text{MIN}\big[1,\exp(-(C_\x(\x) - C_\x(\x'))/\tau ) \big],
\end{align}
rather than the native samples. We found that the acceptance rule over the native samples $z$ works well for training the encoder to propose better initial samples during training. However, because the energy-matching between the native space and the design space is only approximate, we found that increasing the channel depth during validation only increased the Renyi divergence. This is because the ideal distribution is Boltzmann distributed over the design space and so better convergence to an approximate energy-matched Boltzmann distribution in the native space can only increase the Renyi divergence. Even still, the quantum sampler still maintained a lower Renyi divergence in both cases so this does not change the comparative result.

\subsection{Spectral Gap}
\label{app:spectralgap}
To measure the asymptotic convergence of an AQAM with neutral atoms to a Boltzmann distribution, we look at the spectral gaps $\delta \in [0,1]$ of the transition matrices $P_{\z,\z'} = \Gamma_3(\z \vert \z')$ with eigenvalues $\{\lambda\}$; $\delta = 1 - \max_{\lambda \neq 1} \vert \lambda \vert$ generated by the three different update rules. For more details see \cite{Layden2023Quantum-enhancedCarlo}.

\begin{figure}[]
  \begin{center}
    \includegraphics[scale=1.0]{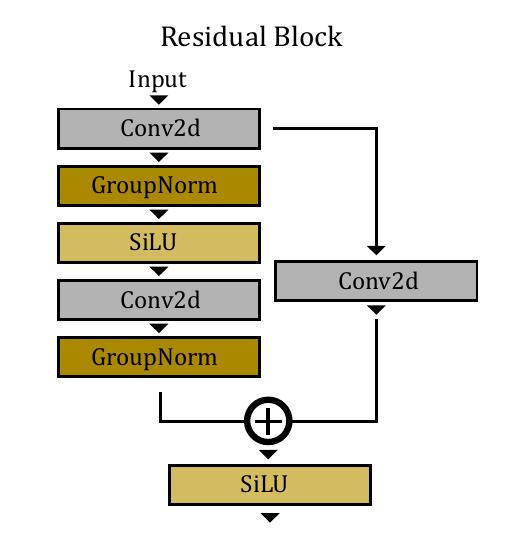}
  \end{center}
  \caption{\textbf{Residual Block (Res. Block):} a core component in the encoder and decoder architectures. The block implements a residual connection between the input and the output of the block, allowing for a more well-bhaved gradient \cite{He2015DeepRecognition}.}
  \label{fig:resblock}
\end{figure}


\section{Adversarial Quantum Autoencoder Model Architecture}
\label{app:architecture}
\begin{figure*}[t]
  \begin{center}
    \includegraphics[scale=1.0]{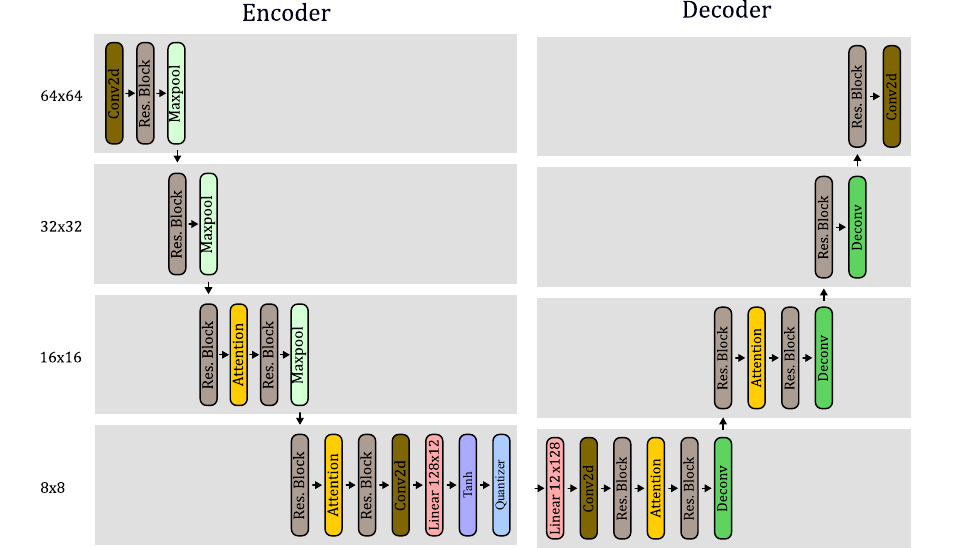}
  \end{center}
  \caption{\textbf{Encoder and Decoder Networks:} the encoder $\tilde q_\ph(\z \vert \x)$ and decoder $ p_\th(\x \vert \z)$ are constructed using a UNet-inspired architecture \cite{Ronneberger2015U-Net:Segmentation} using attention layers  and residual connections \cite{He2015DeepRecognition}.
  The attention layers use the standard attention scaled dot-product \cite{Vaswani2017AttentionNeed} on the $16\times16$ and $8\times8$ levels. Each 2D Convolutional (Conv2d) and residual block (Figure \ref{fig:resblock}) contains one and three convolutions, respectively, with $3\times3$ kernels and +1 padding to maintain the spatial dimensions. Each max pooling layer involves a kernel size of (2x2) and stride of 2 to halve the spatial dimensions. Deconvolutions are performed with a kernel size of (2x2) and stride of 2 to double the spatial dimensions. The architectures use Sigmoid linear unit (SiLU) activation functions. }
  \label{fig:encoder_decoder}
\end{figure*}

The encoder and decoder are implemented using various layers in PyTorch where the parameters $\th$ and $\ph$ are the layer weights.
As outlined in Figure \ref{fig:encoder_decoder}, the autoencoder architecture is based on a convolutional UNet architecture \cite{Ronneberger2015U-Net:Segmentation} where the design resolution is down converted using max pooling layers. Between each down conversion, we use a residual block \cite{He2015DeepRecognition} (Figure \ref{fig:resblock}) to reduce the vanishing gradient and improve model quality.
 
Likewise, for the $16 \times 16$ and $8 \times 8$ pixel embeddings, we also employ attention layers \cite{Vaswani2017AttentionNeed}, where previous layers' outputs are transformed into a query (Q), key (K) of dimension $d_k$, and value matrices (V), to which the attention operation is performed:
\begin{equation}
{\rm Attention}(Q, K, V) = {\rm softmax}(\frac{QK^T}{\sqrt{d_k}})V.
\end{equation}

Once the design is down-sampled to an $8 \times 8$ image representation, it is processed by residual and convolutional blocks before being compressed to a 12-bit spin vector $s_0 \in \{-1, 1\}^{12}$.
The compression first uses a $128 \times 12$ linear layer and clamps the output using a hyperbolic tangent function so the output is in $(-1, 1)$, with a heavy bias toward the boundary values $-1$ and $1$.
Then, the output is quantized using techniques from \cite{Rolfe2017DiscreteAutoencoders, Fajtl2020LatentAutoencoder} to ensure the gradient is calculated correctly on the backward pass.
The spin basis bitstring $s_0 \in \{-1, 1\}^{12}$ is converted to the binary bitstring $\z_0$ using the invertible transformation $s_{0, i} = \frac{\z_{0, i} + 1}{2}$.

For gradient calculations of (\ref{eq:noisy_recon}), we use the spin basis, with the appropriate energy conversions, to ensure the native bitstrings are not zero-biased.
For each time step $t$ for the bitstring sample $\z_t$, we backpropagate on $(\ref{eq:noisy_recon})$ using the time-dependent noise $\epsilon_t$. 
To do so, we generate the noise $\epsilon_t$ using the appropriate sampler and add it as a constant value to $\z_t$.
For each time step, we decode the non-native sample $\x_t \sim p_\theta(\x_t \vert \z_t)$ and backpropagate the loss functions to tune the variational parameters $\th$ and $\ph$. 
The decoder is implemented largely the same way as the encoder, just with the appropriate inverted process. To invert the max pooling layers, we use 2D deconvolutional layers and a $12 \times 128$ linear layer.

\end{document}